\documentclass[preprint,nofootinbib,aps,twocolumn,superscriptaddress,showpacs]{revtex4-1}
\usepackage{color,graphicx,shortvrb,epsfig}

\usepackage{graphicx}
\usepackage{dcolumn}
\usepackage{bm}

\usepackage{colordvi}
\usepackage{epsfig,amssymb,amsmath,latexsym}

\def\lsim{\mathrel{\rlap{\lower4pt\hbox{\hskip1pt$\sim$}}
    \raise1pt\hbox{$<$}}}         
\def\gsim{\mathrel{\rlap{\lower4pt\hbox{\hskip1pt$\sim$}}
    \raise1pt\hbox{$>$}}}         

\def\lsim{\mathrel{\rlap{\lower4pt\hbox{\hskip1pt$\sim$}}
    \raise1pt\hbox{$<$}}}         
\def\gsim{\mathrel{\rlap{\lower4pt\hbox{\hskip1pt$\sim$}}
    \raise1pt\hbox{$>$}}}         

\def\beq{\begin{equation}}
\def\eeq{\end{equation}}
\def\ba{\begin{eqnarray}}
\def\ea{\end{eqnarray}}

\def\<{\langle}
\def\>{\rangle}

\begin{document}

\begin{flushright}
\date{}
\end{flushright}

\title{Partitioning of 1-alkanols in composite raft-like lipid membrane
}
\
\author{Anirban Polley} 
\affiliation{Department of Chemical Engineering, Columbia University, New York City, New York-10027, USA}


\begin{abstract}
Though 1-alkanols are well known to have anesthetic properties, the mode of operation remains enigmatic. We perform extensive atomistic molecular dynamics simulation to study the partitioning of 1-alkanols of different chain-lengths in `raft-like' model membrane made up of mixture of unsaturated dioleoyl-phosphatidylcholine (DOPC) and saturated dipalmitoyl-phosphatidylcholine (DPPC) and cholesterol (Chol) exhibiting phase coexistence of liquid-ordered ($l_o$) - liquid disordered ($l_d$) phase domains. Our simulation shows that the effect of 1-alkanols on the mechanical properties of the membrane has been pronounced with the chain-length of it. In particular, the 1-alkanols prefer to partition in the $l_d$ phase domain. The penetration of the 1-alkanols in the membrane increases significantly for the long-chain alkanol. We also have found the dependency of the order of chains of the lipids and rigidity of the membrane on the 1-alkanols. 
\end{abstract}

\maketitle

\section{\label{sec:level1}Introduction}
\noindent
The phenomena of general anesthesia  \cite{seeman_anesthetic}  is known from a long time and the use of anesthetics in the hospitals are very common for all painless surgical operation. However, the molecular level of understanding in the mechanism of the general anesthesia is not yet understood. It is still debatable whether the general anesthesia is caused by the binding of anesthetics to specific proteins  \cite{Frank_1994,Frank_1997} and block the protein function by changing conformation or it is an indirect lipid-mediated mechanism  \cite{Regen_2009,Ueda_1998,Ueda_2001} while the anesthetics alter the membrane properties such as total volume of the membrane, the volume occupied by the anesthetics within the membrane, the phase transition temperature, the lipid chain order, membrane thickness or the lateral pressure profile of the membrane. 

Apart from anesthesia, the alkanol has been well known as a penetration enhancer for transdermal drug delivery. However, the understanding of the mechanism of the action of penetration enhancer remains opaque. It is interesting to know that 1-alkanol as anesthetics or as penetration enhancer exhibits `cutoff-effect'. The potency of 1-alkanols as anesthetics is increased up to a certain cutoff chain length (dodecanol) \cite{miller_anesthetic} and 1-alkanols with a chain length above cut-off length show no anesthetic potency  \cite{miller_anesthetic}. Similarly, the penetration enhancement property of 1-alkanols increase with increasing the chain length up to decanol and decrease again for 1-alkanol with longer than cut-off chain length. The potency of 1-alkanols as anesthetics and as penetration enhancer both is decreased with the branching of the carbon chains. 

The several experimental studies have been conducted to explore the effect of the anesthetics in the property of the lipid-membrane such as NMR spectroscopic studies that show ethanol molecules having disordering effect on the lipids \cite{Feller_2002,Holte_1997,Barry_1995,Patra_2006,Joaquim_2011,Igor_2012};  X-Ray studies that reveals ethanol having potential to change in the lipid membrane above its main phase transition temperature\cite{Cantor_1997a,Cantor_1997b,Cantor_1998}.
Few molecular dynamics studies also have been performed to study the influence of the anesthetics in the lipid membrane \cite{Patra_2006,Bandyopadhyay_2004,Bandyopadhyay_2006,Smit_2004,
Terama_2008,Griepernau_2007,Faller_2007,
Vierl_1994,Jackson_2007,Gawrisch_1995,Dunn_1998,Mcintosh_1984}.

In the present work, we investigate the effects of 1-alkanols: ethanol, pentanol, and octanol on the symmetric `raft-like model membrane', comprising DOPC, DPPC and Chol, which exhibits phase coexistence of $l_o$-$l_d$ domains.  `Membrane-raft' exhibiting lipid-based compositional heterogeneity has been thought to facilitate  several physiological activities like protein sorting, signaling processes \cite{simons,simons_science2010,simons_nat_rev_2000,anirban_cell15,
TrafficRaoMayor,raftreviews,sharma,debanjan,kripa_2012}. Thus it is important to study the effect of 1-alkanols into the model raft-like membrane. 

The article is arranged as follows: we first discuss the details of the atomistic molecular dynamics simulations of the multicomponent bilayer membrane. Next, we present our main results of the spatial heterogeneity of the components, effects of 1-alkanols on the thickness, penetration, partitioning, order parameter and bending rigidity of the membrane. We end by summarizing our findings.

\section{\label{sec:level2}Methods}
\noindent
{\it Model membrane}\,:\,
 We have studied symmetric 3-component bilayer membrane embedded in an aqueous medium by
atomistic molecular dynamics simulations (MD) using {\it GROMACS-5.1}  \cite{Lindahl}. 
We prepare the bilayer membrane at  $23^{\circ}$C at the relative concentration, $33.3\%$  of DOPC, DPPC, and  Chol, respectively.  
To the symmetric ternary bilayer membrane, we have introduced $25\%$ of 1-alkanols (ethanol, pentanol, and octanol) to water layer from both sides of the symmetric ternary bilayer membrane, respectively. 
All 4 multicomponent bilayer membranes comprise $512$ lipids in each leaflet (with a total $1024$ lipids) and $32768$ water molecules (such that the ratio of water to lipid is $32:1$) to hydrate the bilayer membrane. 
In our simulation study, we choose the compositions of the membrane such that we obtain the bilayer membrane exhibiting $l_o$-$l_d$ phase coexistence \cite{schwille_dopc_dppc_chol}.\\

\noindent
{\it Force fields}\,:\,
The force field parameters of DOPC, DPPC, and Chol are taken from the previously validated united-atom description  \cite{kindt_dopc_dppc_chol,anirban_jpcb12,anirban_cpl13,Tieleman-POPC,mikko}. We have taken the same previously used force-field parameters for the ethanol, pentanol, and octanol  \cite{bockmann_alkanol,anirban_cpl13,MikkoBPJ2006,ramon_jpcb2011,ramon_plosone2013}. In our simulation study, we use improved extended simple point charge (SPC/E) model to simulate water molecules, with an extra average polarization correction to the potential energy function.    
\\

\noindent
{\it Initial configurations}\,:\,
We generate the initial configurations of the symmetric multicomponent bilayer membrane using {\it PACKMOL} \cite{packmol}, where all the components are  homogeneously mixed.  All the symmetric bilayer membrane comprises $342$ DOPC, $340$ DPPC, and $342$ Chol with spatially uniformly randomly placed, and are hydrated with $32768$ water molecules. To the symmetric composite membrane, we introduce $256$ 1-alkanols (ethanol, pentanol, and octanol) uniformly in both sides of the water layers of the bilayer membrane using  {\it PACKMOL} \cite{packmol} as shown in Fig. S1 in the supplementary information (SI).\\

\noindent
{\it Choice of ensembles and equilibration}\,:\,
We simulate the symmetric bilayers for $50$\,ps in the NVT ensemble using a Langevin thermostat to avoid bad contacts arising from steric constraints and then for $500$\,ns in the NPT ensemble ($T = 296$\,K ($23^{\circ}$C), $P =1$\,atm).
To get enough data, we repeat the simulations $4$ times (total $2\mu s$) to study each system. We run the simulations in the NPT ensemble for the first $50$\,ns using Berendsen thermostat and barostat, then using Nose-Hoover thermostat and the Parrinello-Rahman barostat to produce the correct ensemble using a semi-isotropic pressure coupling with the compressibility $4.5\times 10^{-5}$ bar$^{-1}$ for the simulations in the NPT ensemble, long-range electrostatic interactions  by the reaction-field method with cut-off $r_c = 2$\,nm, and the Lennard-Jones interactions using a cut-off of $1$\,nm
\cite{anirban_jpcb12,mikko,patra2004}. We use last $200\,ns$ of the trajectories of the four independent repeated simulations for the data analysis.\\

\noindent
{\it Computation of bending rigidity}\,:\, 
To calculate bending rigidity, we have employed spectral method, where we need to calculate structure factor of the fluctuation of the lipid head groups. In this method, we need to use a large membrane, otherwise, we would get very noisy data. Therefore, we have carried out the simulation of $9216$ lipid membrane  size $\approx 48 \times 48$\,nm using the already equilibrated last configuration of $1024$ lipid patch of the $500$\,ns simulation. Here, we translate the $1024$-lipid membrane patch with 9 positions like a $3\times3$ matrix-form using vmd, such that we generate the initial configuration of a large bilayer membrane of $9 \times 1024$ lipids for the further simulation. We run large bilayer membrane for $100$\,ns in NPT ensemble using Nose-Hoover thermostat and the Parrinello-Rahman barostat to produce the correct ensemble using a semi-isotropic pressure coupling with the compressibility $4.5\times 10^{-5}$ bar$^{-1}$ as described above. After the simulation, we collect the position of the center of mass of the head group of lipids and use to calculate the structure factor.

\section{\label{sec:level3}Results and Discussion}

To confirm that we attend thermally and chemically equilibrated bilayer membrane, we calculate the time evolution of the total energy of the system and area per lipid and when the values attend its asymptotic behavior with a very small fluctuation, we can be satisfied as we obtain equilibrated bilayer membrane \cite{anirban_jcb14,anirban_jpcb12}.

\begin{figure*}[h!t]
\begin{center}
\includegraphics[width=18.0cm]{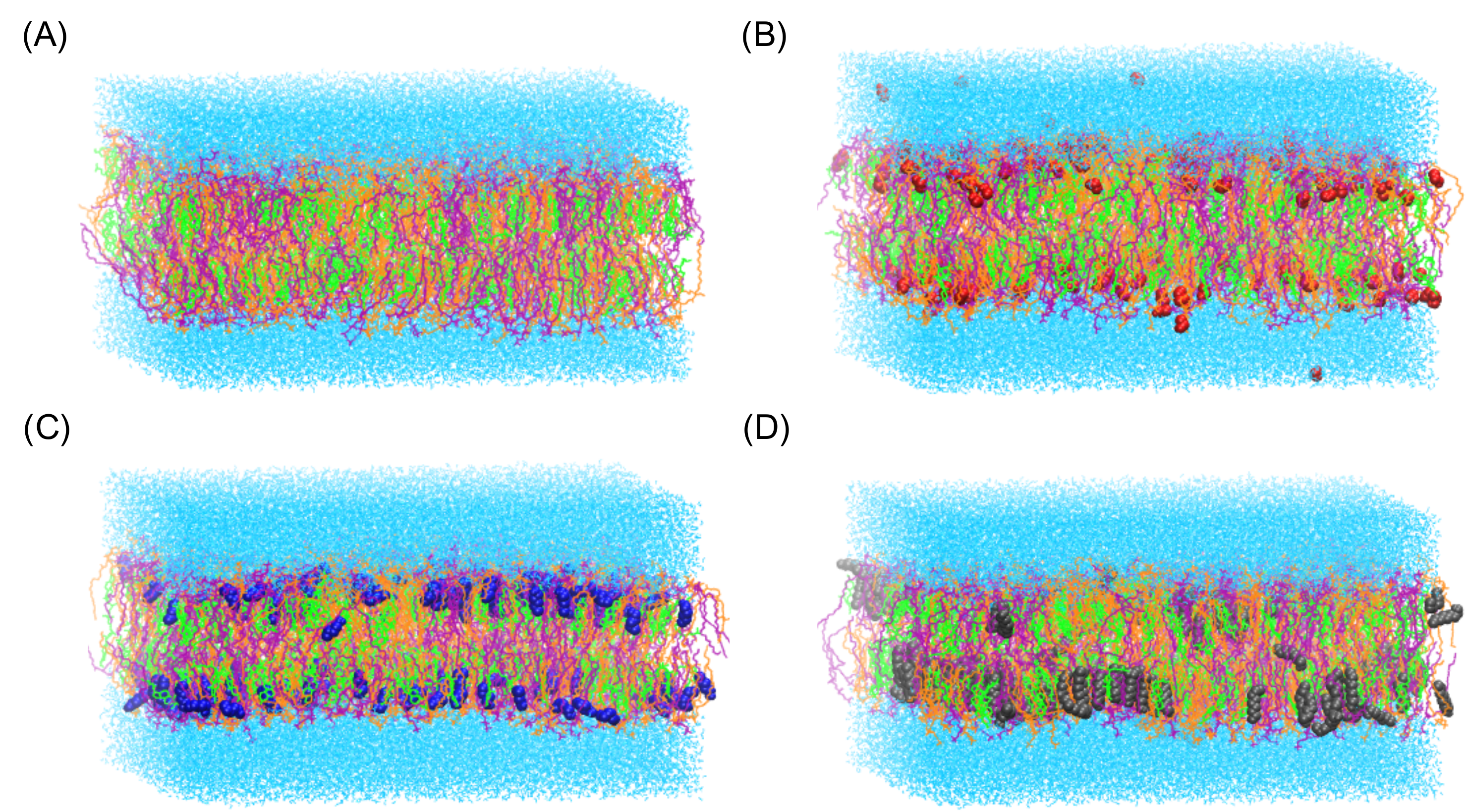}
\caption{The snapshot of symmetric composite bilayer membrane comprising DOPC (purple), DPPC (orange), Chol (green), water (cyan) (A) without 1-alkanol, (B) with Ethanol (ETH) (red), (C) Pentanol (PCT) (blue) and (D) Octanol (OCT) (gray) are shown  respectively.
}
\label{snapshot}
\end{center}
\end{figure*}

We start our simulation from the initial configuration, where lipid components are homogeneously mixed and the 1-alkanols are in the aqueous layers of both sides of the bilayer membrane shown Fig. S1 in SI. After the long simulation, we get the equilibrated patch of bilayer membrane and the last frame of the simulation have been shown in Fig. \ref{snapshot}, where all the components, especially 1-alkanols are highlighted to show their preferable arrangement in the composite bilayer membrane.

\subsection{Transverse heterogeneity of the components in membrane}
\begin{figure*}[h!t]
\begin{center}
\includegraphics[width=18.0cm]{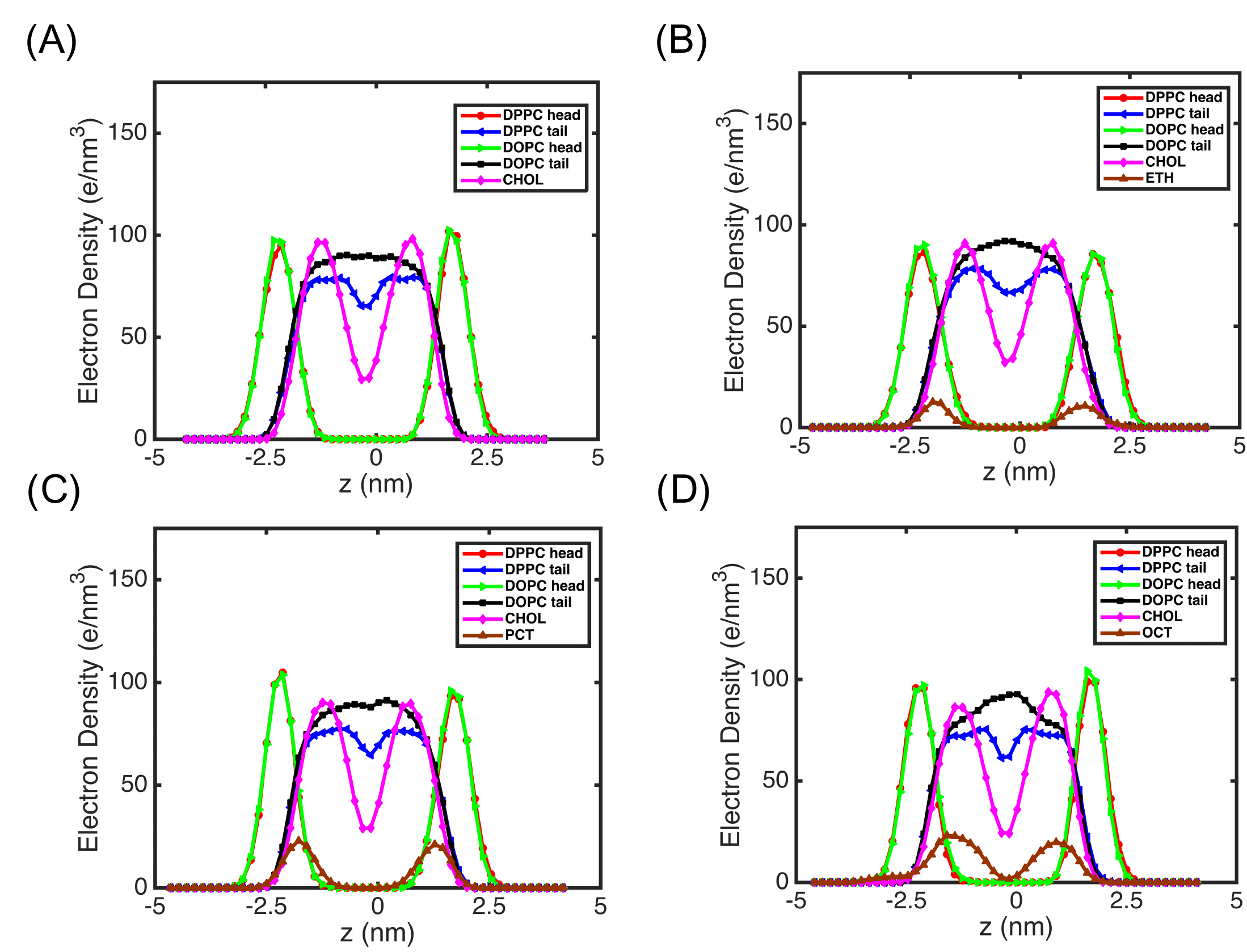}
\caption{The electron density of each component vs. z-axis (normal to membrane) of the symmetric composite bilayer membrane (A) without 1-alkanol, (B) with Ethanol (ETH), (C) Pentanol (PCT) and (D) Octanol (OCT) are shown in  respectively.
}
\label{ed_lat}
\end{center}
\end{figure*}

The electron density profile of components in the membrane without and with 1-alkanols has been shown in Fig. \ref{ed_lat}. Here, the electron density of the head groups and tails of DPPC, DOPC lipids and Chol and 1-alkanols, respectively in the z-axis are plotted (In our simulation, z-axis is representing the normal to the surface of the membrane lying in the xy-plane). 

We find the peaks of the head-group of the lipids (DPPC and DOPC) are pronounced averaged at 
$2.058$ \,nm, $1.915$\,nm, $1.887$\,nm and $1.859$\,nm 
respectively for without and with 1-alkanols (ethanol, pentanol, and octanols). (This is the averaged value of the thickness of the one leaflet of the membranes.)

\begin{figure*}[h!t]
\begin{center}
\includegraphics[width=18.0cm]{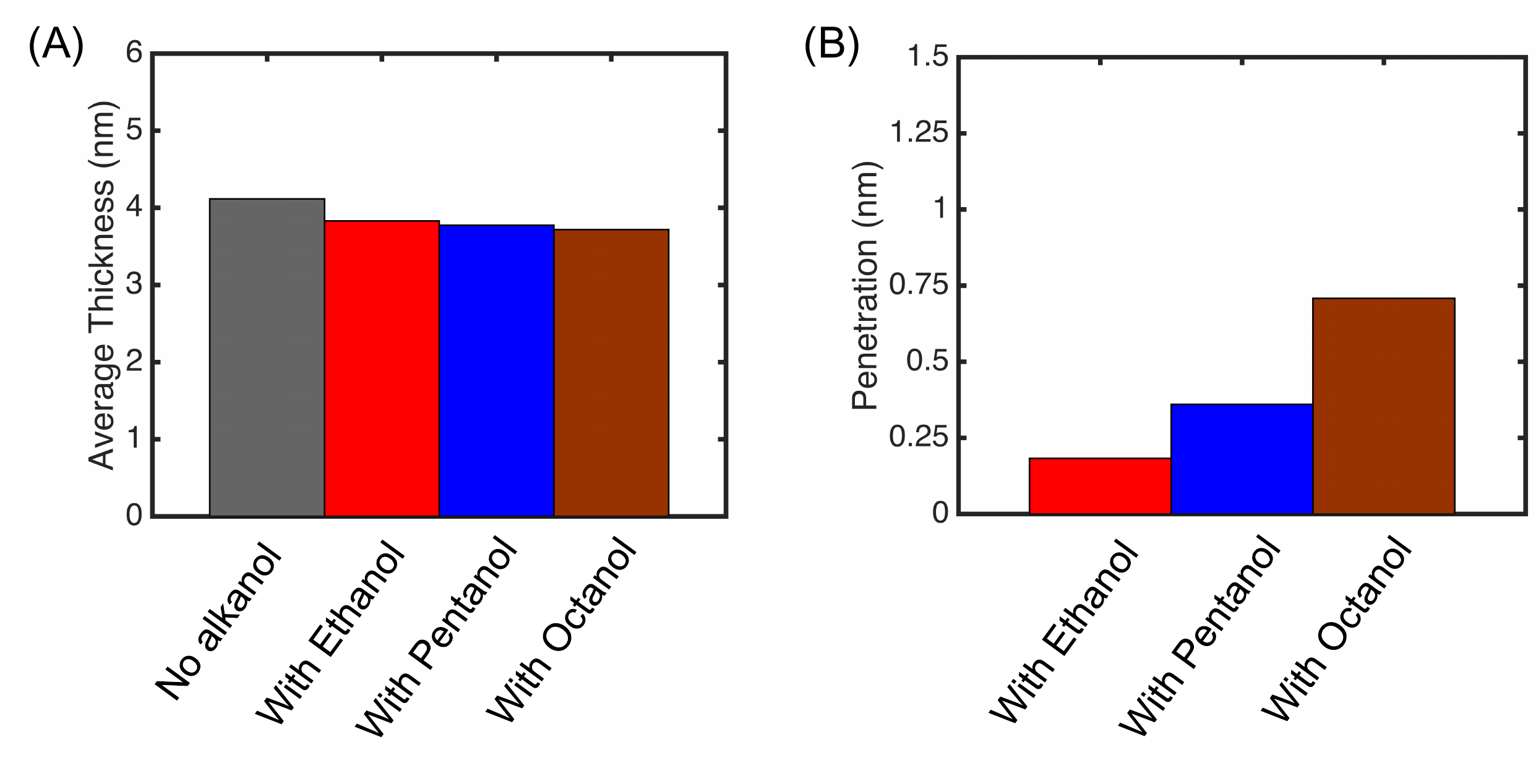}
\caption{(A) The average thickness of the membrane without 1-alkanol and ethanol (chain length =$2$), pentanol (chain length =$5$) and octanol (chain length =$8$) and (B) the penetration of ethanol, pentanol and octanol are shown in respectively.
}
\label{th_pene}
\end{center}
\end{figure*}

From the two peaks of the head-group the lipids (DPPC and DOPC), we can measure the average thickness of the membranes with and without 1-alkanols. 
Here, we find that the average thickness of the lipid membrane has been decreased with the number of carbon chain length in 1-alkanol as the carbon chain length of ethanol, pentanol, and octanol are $2$, $5$ and $8$, respectively as shown in Fig. \ref{th_pene} (a).

Again, we find from the electron density profile of the 1-alkanols highlighted in brown shown in Fig. \ref{ed_lat}, that the peaks of the electron density of 1-alkanols are closest (average in two leaflets $1.1509$\,nm) to 0 (center of the membrane, $z=0nm$) for octanol and farthest for ethanol (average in two leaflets $1.732$\,nm), and in between value (average in two leaflets $1.527$\,nm) for pentanol, which is related to the depth of penetration of 1-alkanols at equilibrium. To illustrate the depth penetration into the membrane, we plot the mean depth of penetration for 1-alkanols with increasing chain length as shown in Fig. \ref{th_pene} (b), which shows that the depth of penetration of 1-alkanols in the membrane increases with the increase of carbon chain length of the 1-alkanols.

\subsection{Lateral heterogeneity of the components in membrane}

To elucidate the spatial heterogeneity of the components, especially 1-alkanols, we calculate the spatial number density of each component of the membrane. We collect the last $200$\,ns trajectories of the simulations of each system and calculate the center of mass of each component and project it on the xy-plane and calculate the number density by binning the data with bin-size $0.4$\,nm.

\begin{figure*}[h!t]
\begin{center}
\includegraphics[width=18.0cm]{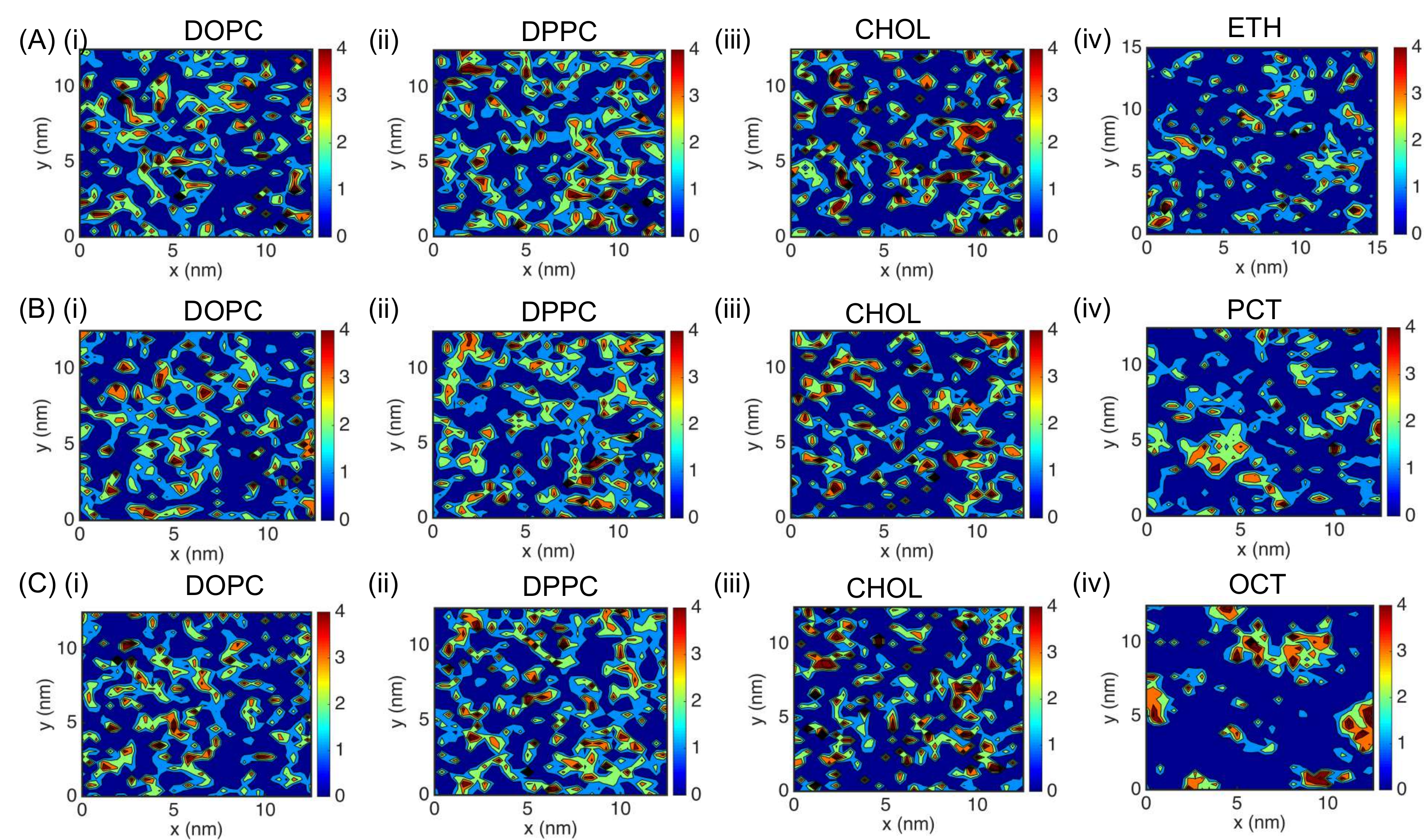}
\caption{The spatial number density of each component in the membrane with ethanol (ETH), pentanol (PCT) and octanol (OCT) are shown in 
(A), (B) and (C), respectively\cite{anirban_jpcb12,anirban_jcb14}. 
}
\label{num_density}
\end{center}
\end{figure*}

The spatial number density of each component of the bilayers with ethanol, pentanol, and octanol are shown in Figure \ref{num_density} (A), (B) and (C), respectively. We find that all the symmetric membrane exhibits phase coexistence of DPPC-rich $l_o$ domain and DOPC-rich $l_d$ domains, while Chol is also accumulated in DPPC-rich domains. Again, we find that ethanol, pentanol and octanol are partitioning in the DOPC-rich domains shown in Fig. \ref{num_density}.

\begin{figure*}[h!t]
\begin{center}
\includegraphics[width=18.0cm]{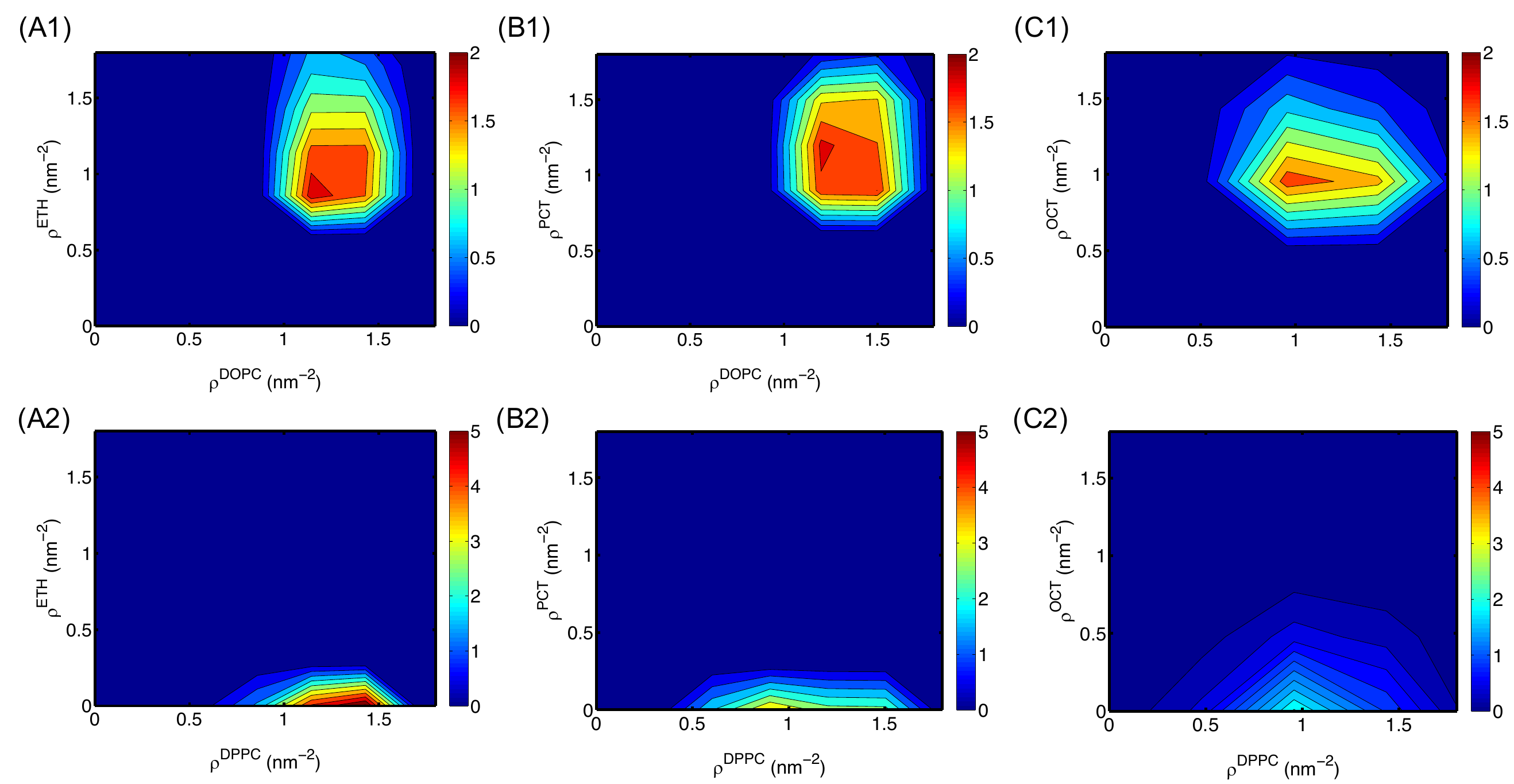}
\caption{The joint probability distribution between (A1) ethanol (ETH) and DOPC, (A2) ETH and DPPC, (B1) pentanol (PCT) and DOPC, (B2) PCT and DPPC, 
(C1) octanol (OCT) and DOPC and (C2) OCT and DPPC are shown respectively \cite{anirban_jpcb12}. 
}
\label{JPD}
\end{center}
\end{figure*}

In Fig. \ref{JPD}, the joint probability distribution (JPD) shows that a distinct peak along the diagonal for the JPD of DOPC to all three 1-alkanols, while   there is no diagonal peak in JPD for any 1-alkanols to the DPPC-rich domain.

To extend our the investigation on the spatial partitioning of the anesthetics, we define the `correlation coefficient' from the normalized cross-correlation  \\
between the 1-alkanols solutes to the DOPC/DPPC-rich domain \cite{anirban_jcb14,anirban_cell15} as 
\begin{widetext}
$C(\rho^{DOPC/DPPC}(r),\rho^{1alkanol}(r))= \\ \nonumber
\frac{\langle \rho^{DOPC/DPPC}(r) \rho^{1-alkanol}(r) \rangle - \langle \rho^{DOPC/DPPC}(r) \rangle \langle \rho^{1alkanol}(r) \rangle}{\sqrt{\langle \rho^{DOPC/DPPC}(r)^2 \rangle - \langle \rho^{1alkanol}(r) \rangle^2} \sqrt{\langle \rho^{DOPC/DPPC}(r)^2 \rangle - \langle \rho^{1alkanol}(r) \rangle^2}} $
\end{widetext}
averaged over space (denoted by $C_{uu}$), while $\rho^{DOPC/DPPC}(r)$ and $\rho^{1-alkanol}(r)$ are the spatial number densities of the DOPC/DPPC and 1-alkanol, respectively with the same positional coordinate $r(x,y)$.

\begin{figure*}[h!t]
\begin{center}
\includegraphics[width=9.0cm]{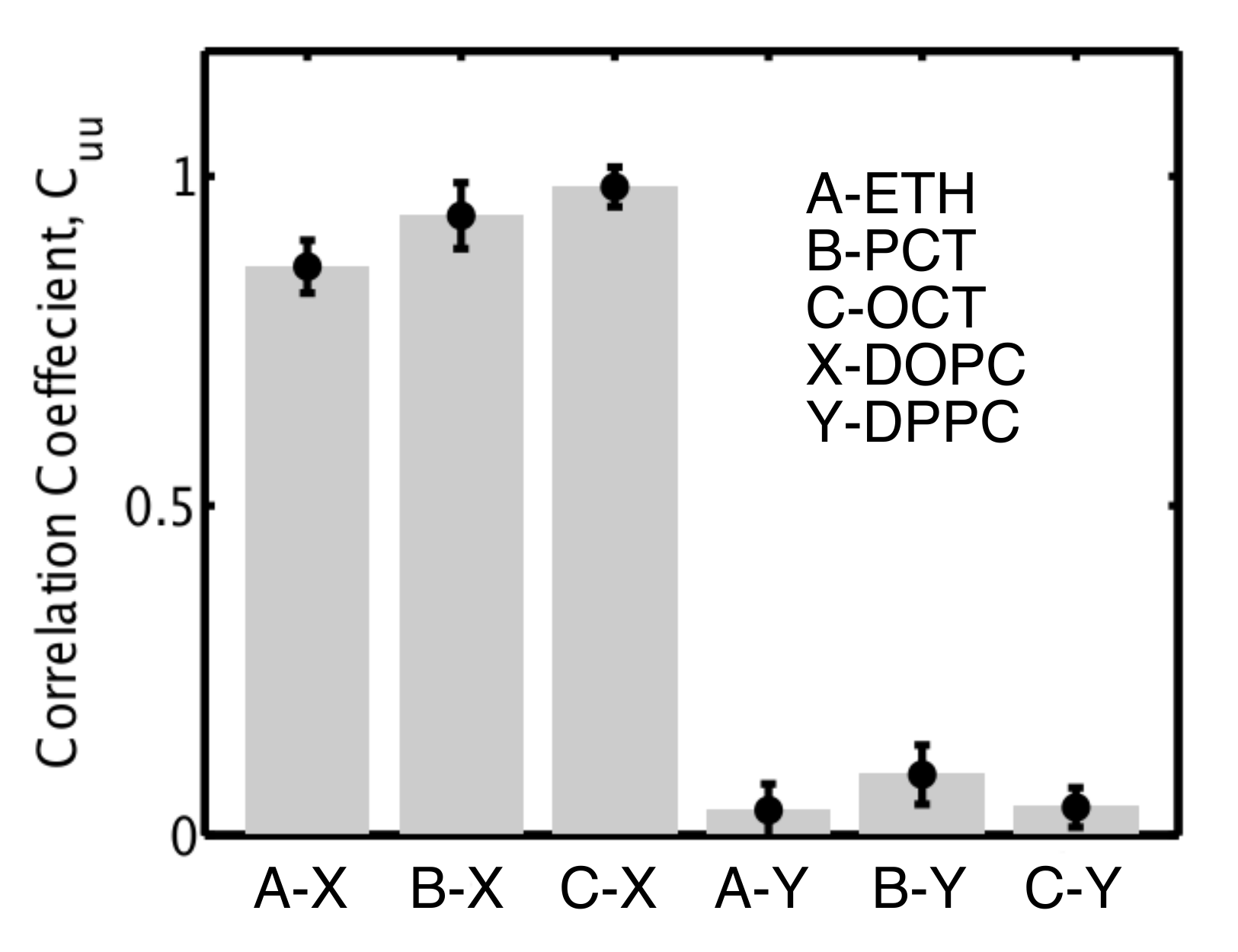}
\caption{The cross-correlation between ethanol (ETH) and DOPC (A-X), ETH and DPPC (A-Y),  pentanol (PCT) and DOPC (B-X), PCT and DPPC (B-Y), octanol (OCT) and DOPC (C-X), and OCT and DPPC (C-Y) are shown respectively\cite{anirban_cell15,anirban_jcb14}. 
}
\label{partition}
\end{center}
\end{figure*}

We compute $C_{uu}$ for DOPC to ethanol, pentanol, and octanol and PSM to ethanol, pentanol, and octanol, respectively bilayer membranes shown in Figure \ref{partition} which indicates that 
the values of $C_{uu}$ are high for DOPC-rich domain to
 ethanol/ pentanols/ octanols-rich domains while that values are low for DPPC to all 1-alkanols. This suggests that all three 1-alkanols (ethanol, pentanol, and octanol) prefer to partition into the DOPC-rich domain, while 1-alkanols do not partition in DPPC-Chol rich domain.

\begin{figure*}[h!t]
\begin{center}
\includegraphics[width=18.0cm]{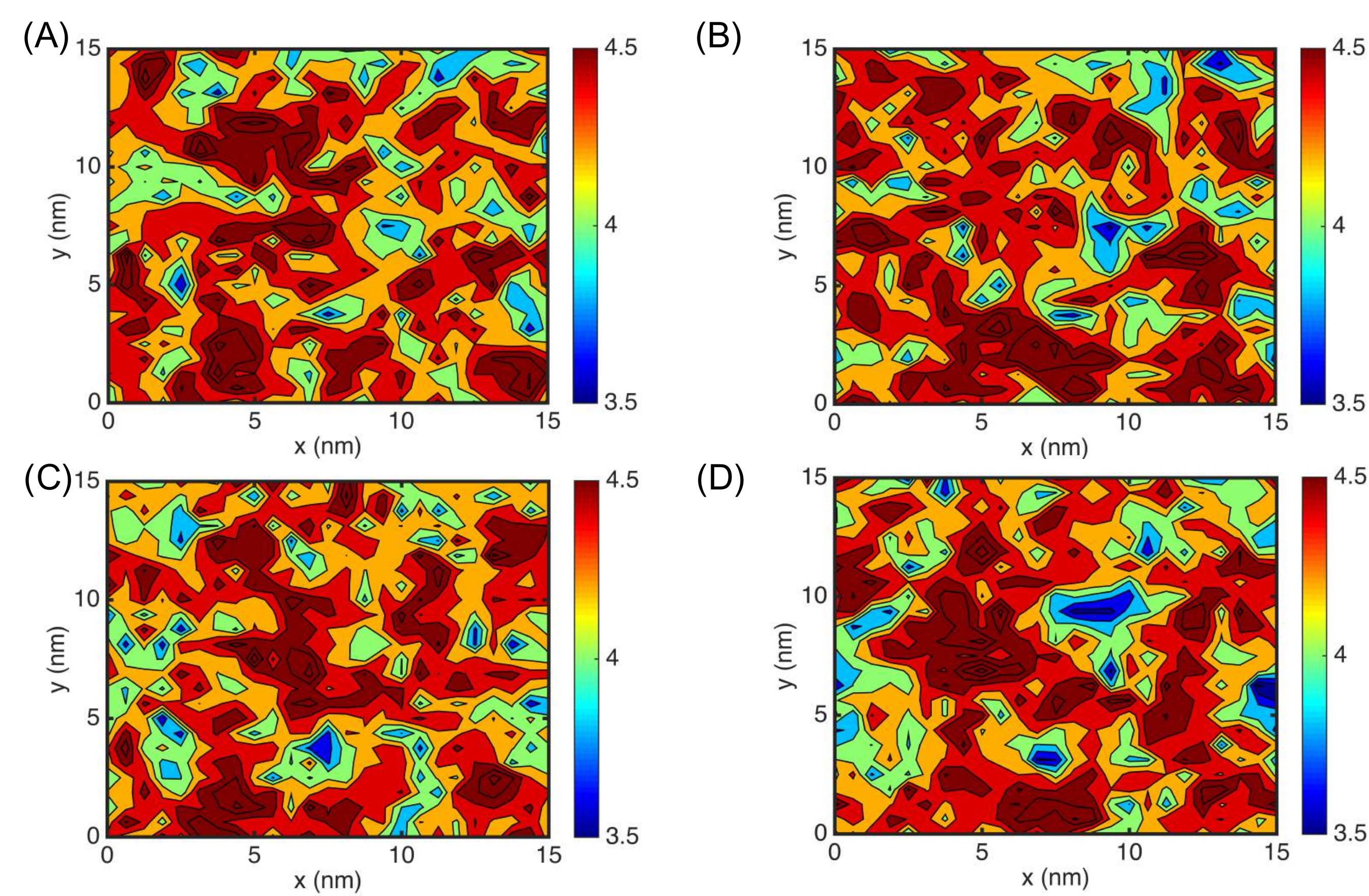}
\caption{The spatial thickness of composite bilayer membrane (A) without 1-alkanol, (B) with ethanol (ETH), (C) with pentanol (PCT) and (D) with octanol (OCT) are shown respectively \cite{anirban_jpcb12}. 
}
\label{spatial_th}
\end{center}
\end{figure*}

\subsection{Spatial heterogeneity of thickness}
Here, we compute the spatial  heterogeneity of thickness for all membranes. To compute the local thickness of the membrane, we first collect the position of P-atom of both DPPC and DOPC lipids of both upper and lower leaflets. Now, we divide the xy-plane in both leaflets into many small boxes with grid size $0.4$\,nm to calculate the mean coordinate of each grid in each leaflet from the collected data of positions of P-atoms. The difference between the $z$-coordinate corresponding to two layers with same (x,y) -coordinate gives us the thicknss of the membrane, $d$ at that local coordinate (x,y). Thus, we calculate the spatial thickness heterogeneity of the membrane.

Figure \ref{spatial_th} shows that the phase coexistence of thicker $l_o$ and thinner $l_d$ phase domains are exhibited in all bilayer membranes.

\subsection{Effects of 1-alkanols on the order parameter}
\begin{figure*}[h!t]
\includegraphics[width=18.0cm]{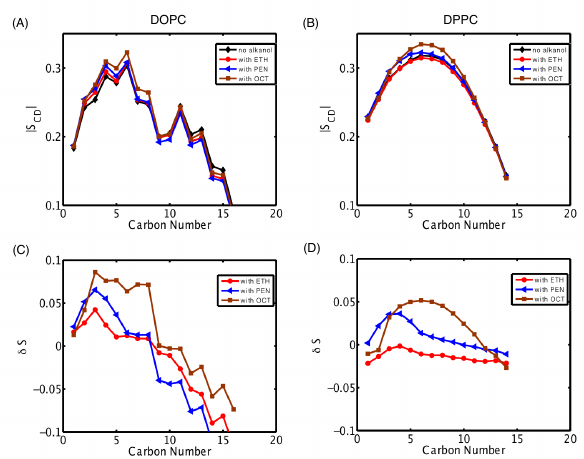}
\caption{(A) the deuterium order parameter (S) vs. carbon number of the acyl chain of the lipids of the multicomponent symmetric membrane without (black) and with ethanol (red), chloroform (blue), and methanol(green), respectively. 
(B) the change of deuterium order parameter ($\delta s$) vs. carbon number of the acyl chain of the lipids in the membrane with anesthetics compared to without anesthetic lipid membrane.  
}
\label{scd}
\end{figure*}

We calculate the deuterium order parameter, $S$ of the acyl chains of the DOPC and DPPC lipids in the membrane respectively, which is defined as $S=\langle \frac{3}{2}(cos^{2}\theta)-\frac{1}{2} \rangle$ where $\theta$ is the angle between the carbon atom - hydrogen (deuterium) atom and the bilayer normal shown in Fig. \ref{scd} (a) and (b). Here, we find two important things: one is that the value of the $|S|$ for DPPC is more than that of DOPC, which suggests that DPPC-rich domain is more ordered ($l_o$) phase, while DOPC-rich domain is less ordered ($l_d$) phase, where acyl chains of the lipids are more floppy. Secondly, we find that the value of  $S$ is affected most in octanol, the value of that is changed lowest in ethanol.

To illustrate the effect of the 1-alkanols on $S$, we define, \\
$\delta S=\frac{S^{A}-S^{0}}{S^{0}}$ where $S^{A}$ and $S^{0}$ are the order parameter of the lipid chains of the symmetric multicomponent bilayer membrane {\em with} and  {\em without} ethanol, pentanol and octanol, respectively and we find that change of the order parameter of the lipids is highest in presence of octanol, while the change of that is lowest in ethanol.

\subsection{Orientation of O-H and C1-Cn bonds}

We characterize the orientation of the vector (OH) between O and H -atoms and that another vector (C1Cn) between the coordinate of first C-atom (C1) and last C-atom (Cn),  where the value of $n=2$, $5$ and $8$ for ethanol, pentanol and octanol, respectively. To quantify the orientation of the two vectors: OH and C1Cn, we define an orientational order parameter, $I_{\beta}$ as\, \\
$I_{\beta}=\langle Cos(\beta)\rangle$,
 where $\beta$ is the angle between the vector OH/C1Cn and the normal (z-axis) of the surface of the membrane.
 
\begin{figure*}[h!t]
\begin{center}
\includegraphics[width=18.0cm]{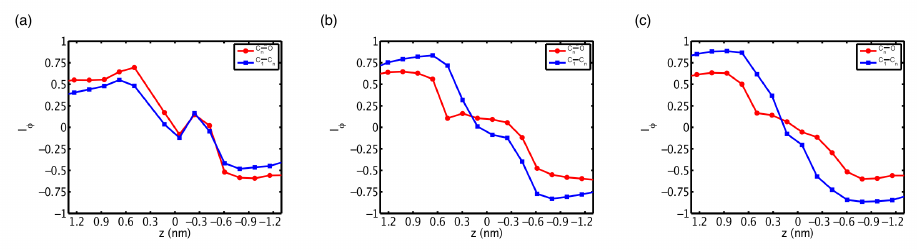}
\caption{Orientation of O-H and C1-Cn vectors of the ethanol, pentanol, and octanol are shown in Fig. \ref{orientation}, respectively.  
}
\label{orientation}
\end{center}
\end{figure*}

 Fig. \ref{orientation} shows the orientational order parameter,  $I_{\beta}$ for the ethanol, pentanol, and octanol, respectively. We find that the value of $I_{\beta}$ close to $\pm 1$ near to the head groups of the lipid membrane and the value of it are zero and switch its value across the center of the membrane. The value of  $I_{\beta}$ of the OH-vector for ethanol is greater than that of C1Cn vector, while the value of OH-vector is less than that of C1Cn-vector for octanol. 
 
 Therefore, Fig. \ref{orientation} suggests that the value of orientation order parameter for C1Cn increases with the acyl-chain length as the acyl chain of the 1-alkanol arranged in more ordered fashion with the increase in chain length.

\subsection{Effects of 1-alkanols on bending modulus and elasticity}

\begin{figure*}[h!t]
\begin{center}
\includegraphics[width=18.0cm]{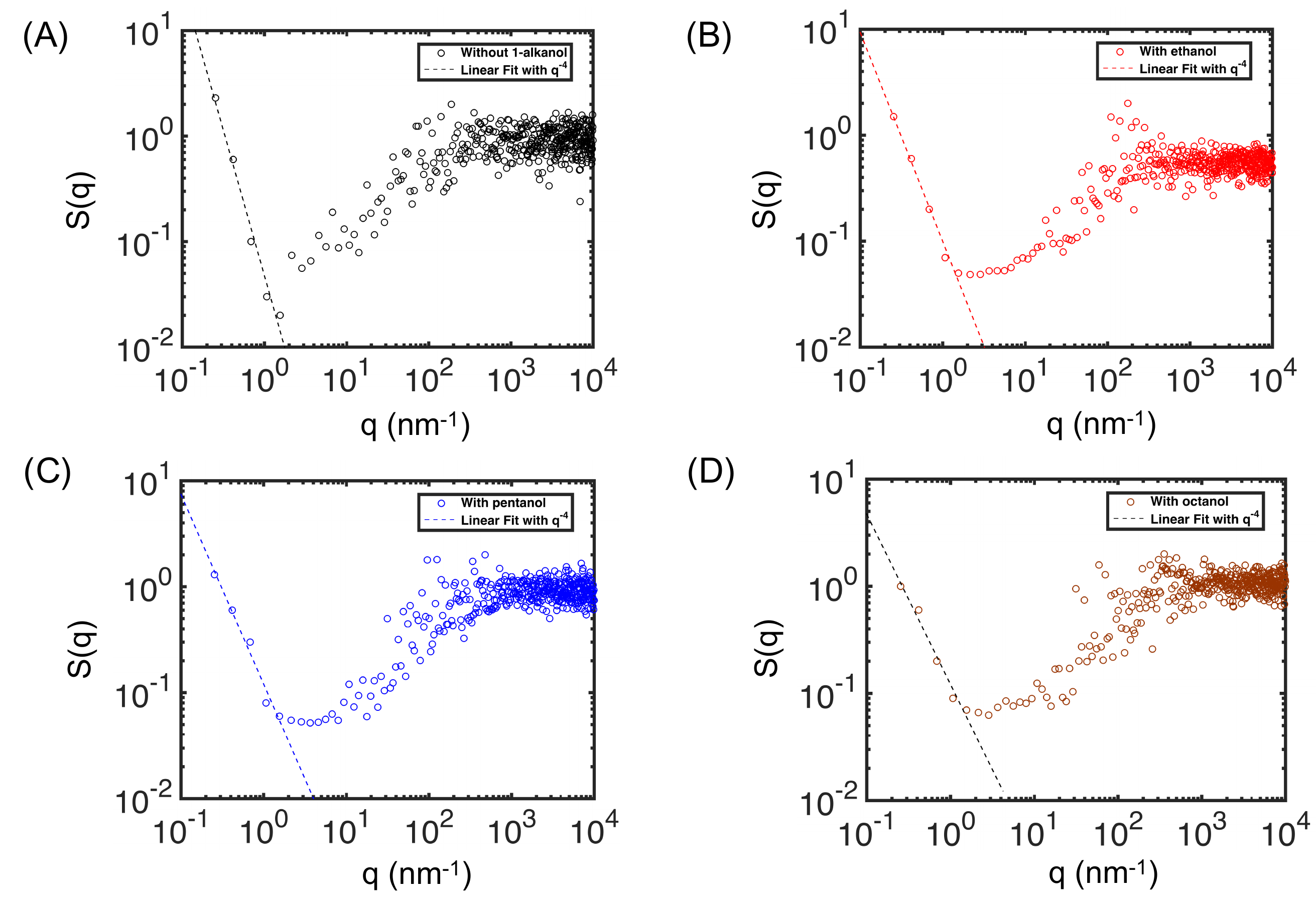}
\caption{The fluctuation spectra S(q) versus wave vector q for the membrane (A) without 1-alkanol, (B) with ethanol, (C) with pentanol and (D) with octanol, respectively are shown. Here, the bending rigidity is calculated using the long wavelength undulation spectrum. The fitting is carried out using the theoretical $q^{-4}$ line and is denoted as the dashed line.  
}
\label{rigidity}
\end{center}
\end{figure*}

If we assume our bilayer as a flat surface $u(x,y)$ with fluctuation over minimum surface energy at $u(x,y)=u_0$ and the energy costs due to the thermal fluctuation, then we can write the Hamiltonian of the membrane as 

\begin{eqnarray} \nonumber 
H[u(x,y)]&=&\frac{1}{2}\int(\kappa | \nabla^2 u(x,y) |^2  \nonumber \\
&&+ \gamma |\nabla u(x,y) |^2)dx dy
\label{Hamiltonian}
\end{eqnarray}

Here, the first term in the Hamiltonian is the energy cost due to the bending of the membrane with a constant $\kappa$, the bending rigidity whereas the second term corresponds to the area fluctuation of the surface with constant $\gamma$, the surface tension of the membrane. 

If we transform the Eqn.\ref{Hamiltonian} into Fourier Space by 
$u(\vec{r})=\sum\limits_{\vec{q}} u(\vec{q}) e^{i \vec{q} \cdot \vec{r}}$ where, $\vec{r}\equiv(x,y)$ in two dimensional real space and $\vec{q}\equiv(q_x,q_y)$ in two dimensional Fourier space, then we get by the use of equipartition theorem,

\begin{equation}
S(q)=\frac{K_B T}{A_L(\kappa q^4 + \gamma q^2)}
\label{structure_factor}
\end{equation}

where, $S(q) \equiv \frac{1}{2}N<|u(q)|^2>$ is the structure factor of the membrane surface $u(x,y)$ with projected area per lipid $A_h=A_L/N$ of $N$ number of lipids in each leaflet with total area $A_L$.

As the surface tension $\gamma$ for the lipid bilayer membrane is approximately zero ($\gamma \approx 0$), the Eqn.\ref{structure_factor} turns into
\begin{equation}
S(q)=\frac{K_B T}{A_L \kappa q^4}
\label{structure-factor-approx}
\end{equation}

However, the Eqn.\ref{structure_factor} tells us that the nonzero surface tension $\gamma$ affects the fluctuation spectra $S(q)$ significantly ($S(q)\propto q^{-2}$ for $q\ll\sqrt{\frac{\gamma}{\kappa}}$).
 
 Fig. \ref{rigidity} shows the structure factor $S(q)$ vs. $q$, which is fitted to the straight line for small wave vector, $q$ (See {\it Section: Methods}).

\begin{table*}[h!t]  
\begin{center}
\begin{tabular}{|c|c|}   \hline
System                    &   $\kappa$  ($k_{B}T$)     \\  \hline
DOPC,DPPC,CHOL                        &            $50.3857\pm 0.9590$         \\  \hline
DOPC,DPPC,CHOL and ethanol               &         $33.5272\pm 0.6002$            \\  \hline
DOPC,DPPC,CHOL and pentanol            &       $29.3434\pm 0.5157$          \\  \hline
DOPC,DPPC,CHOL and octanol              &      $22.9672\pm 0.7419$       \\  \hline
\end{tabular} 
\label{bending_rigidity}
\caption{The value of the $\kappa$ of the multicomponent bilayer membrane without and with 1-alkanols are given in Table-I.
}
\end{center}
\end{table*}

 From the slope of the Fig. \ref{rigidity}, we calculate the bending rigidity of the membrane without and with ethanol, pentanol, and octanol, respectively using equation \ref{structure-factor-approx} and are tabulated in Table-I \cite{george_bending_jctc13}. It suggests that  the rigidity of the membrane decrease with the acyl-chain length of 1-alkanols.

\section{Conclusion}
We present the effect of 1-alkanols with different chain-length in the symmetric `raft membrane' comprising DOPC, DPPC and Chol using atomistic molecular dynamics simulation. Here, we explore the consequence of 1-alkanols by computing the physical properties of the membrane. Our main findings in the present study are: (i) the thickness of the membrane decreases in presence of 1-alkanol and it's value decrease with the increase of the chain length of the 1-alkanols. (ii) the penetration of the 1-alkanols increases with length of chain length of it. (iii) 1-alkanols prefer to partition in the DOPC-rich $l_d$ domain in the 'raft-membrane'. (iv) The deuterium order parameter of the acyl chain of the lipids are decreased in presence of 1-alkanols and its value decreases with the chain length of 1-alkanols. (v) With the increase of chain-length in the 1-alkanols, the orientation order parameter of the acyl chain of 1-alkanols increases. (vi) The value of orientation order parameter for acyl chain (C1Cn) increases with the chain length of acyl chain of the 1-alkanol. (vii) The bending rigidity of the membrane decreases in presence of 1-alkanols.

 Our study shows that the mechanical properties of the membrane change significantly in presence of 1-alkanols, which partition in the $l_d$ domain and the effect on the membrane is pronounced with chain-length of 1-alkanol molecules. Our work would be useful for the study of 1-alkanols and its anesthetic effect.
  
\section{Acknowledgement}
AP  gratefully acknowledges the hospitality of generous the computing facilities of `RRIHPC' clusters at RRI, Bangalore, India and Columbia University, New York. The author also thanks the support of 
TCSC - Tampere Center for Scientific Computing resources, Finland.




\bibliographystyle{apsrev4-1}


%

\end{document}



\makeatletter 
\def\subsubsection{\@startsection{subsubsection}{3}{10pt}{-1.25ex plus -1ex minus -.1ex}{0ex plus 0ex}{\normalsize\bf}} 
\def\paragraph{\@startsection{paragraph}{4}{10pt}{-1.25ex plus -1ex minus -.1ex}{0ex plus 0ex}{\normalsize\textit}} 
\renewcommand\@biblabel[1]{#1}            
\renewcommand\@makefntext[1]%
{\noindent\makebox[0pt][r]{\@thefnmark\,}#1}
\makeatother 
\renewcommand{\figurename}{\small{Fig.}~}
\sectionfont{\large}
\subsectionfont{\normalsize} 



\def\be{\begin{equation}}
\def\ee{\end{equation}}
\def\bea{\begin{eqnarray}}
\def\eea{\end{eqnarray}}
\def\a{\alpha}
\def\b{\beta}
\def\d{\nabla}
\def\p{\partial} 
\def\n{{\bf n}}
\def\t{\tau}
\def\tv{\tau_v}
\def\ta{\tau_a}
\def\td{\tau_d}
\def\v{{\bf v}}
\def\x{{\bf x}}
\def\nn{\nonumber}
\def\sa{\sum_{\alpha}}
\def\sab{\sum_{\alpha \beta}}
\def\si{\sum_{i}}
\def\sij{\sum_{ij}}
\def\la{\langle}
\def\ra{\rangle}
\def\l{\left(}
\def\r{\right)}
\def\e{\epsilon}

\newcommand{\affA}{Raman Research Institute, C.V. Raman Avenue, Bangalore 560080, India}
\newcommand{\affB}{National Centre for Biological Sciences (TIFR), Bellary Road, Bangalore 560065, India}
\newcommand{\affC}{Tampere University of Technology, Korkeakoulunkatu 10, 33720 Tampere, Finland}

\title{}
\begin{center}
{\Large {\bf Supplementary Information}}
\vskip 0.5 cm
{\large {\bf Partitioning of 1-alkanols in composite raft-like lipid membrane }}
\vskip 0.2cm
{\large Anirban Polley}
\end{center}

\vskip 0.5cm

\renewcommand{\thepage}{S\arabic{page}}

\renewcommand{\thefigure}{S\arabic{figure}}

\begin{figure*}[h!t]
\begin{center}
\includegraphics[width=18.0cm]{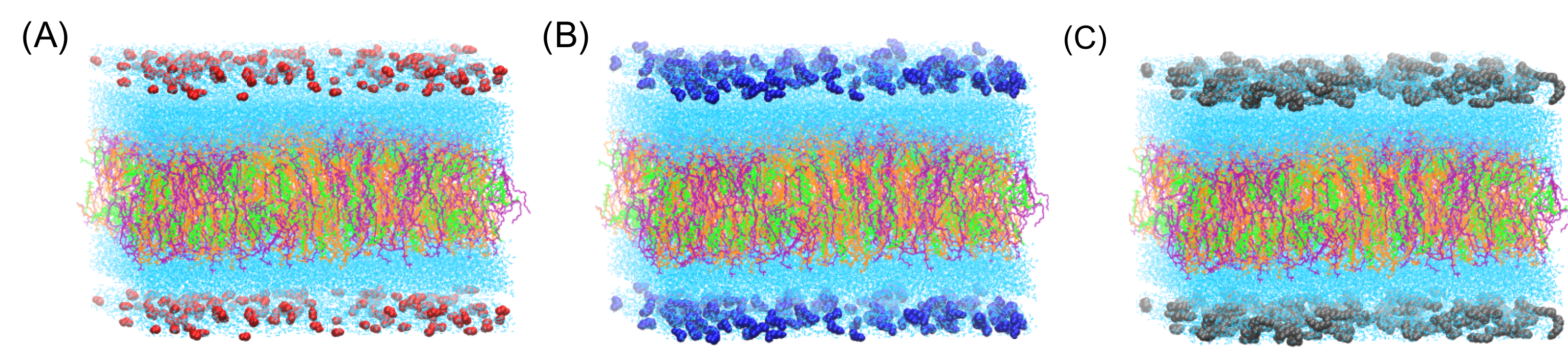}
\caption{ Snapshot of initial configuration of the symmetric bilayer membranes comprising DOPC (purple), DPPC (orange), Chol (green), water (cyan) (A) without 1-alkanol, (B) with Ethanol (ETH) (red), (C) Pentanol (PCT) (blue) and (D) Octanol (OCT) (gray) are shown  respectively. 
}
\label{S1}
\end{center}
\end{figure*}